\documentstyle[12pt]{article}

\begin{document}

\setcounter{page}{1}

\pagestyle{plain} \vspace{1cm}
\begin{center}
\Large{\bf Chaotic inflation on the Randall-Sundrum 2-brane model}\\
\small \vspace{1cm} {\bf Kourosh Nozari}\quad and \quad{\bf Siamak Akhshabi} \\
\vspace{0.5cm} {\it Department of Physics,
Faculty of Basic Sciences,\\
University of Mazandaran,\\
P. O. Box 47416-95447, Babolsar, IRAN\\
knozari@umz.ac.ir\\
s.akhshabi@umz.ac.ir}

\end{center}
\vspace{1.5cm}
\begin{abstract}
We construct an inflation model on the Randall-Sundrum I (RSI) brane
where a bulk scalar field stabilizes the inter-brane separation. We
study impact of the bulk scalar field on the inflationary dynamics
on the brane. We proceed in two different approaches: in the first
approach, the stabilizing field potential is directly appeared in
the Friedmann equation and the resulting scenario is effectively a
two-field inflation. In the second approach the stabilization
mechanism is considered in the context of a warp factor so that
there is just one field present that plays the roles of both
inflaton and stabilizer. We study constraints imposed on the model
parameters from recent observations.
\end{abstract}
\vspace{2cm}
\newpage
\section{Introduction}
The idea of braneworld models is an exciting way to solve the
problems facing standard cosmology and particle physics. The basic
idea is that the standard model matter is confined to a $3$-brane
while gravity propagates in the higher dimensional bulk. This means
that gravity is fundamentally a higher dimensional interaction and
we only see the effective $4D$ theory on the brane. One of the first
suggestions in this respect was the model proposed by Arkani-Hamed
\emph{et al} [1]. In this model the difference between the
fundamental scales of gravity and electroweak is due to the
existence of large extra dimensions that are accessible only for
graviton and possibly non-standard matter.  Later, a different setup
was proposed by Randall and Sundrum [2]. Their model has two
Minkowski branes and a single extra dimension. The branes have equal
and opposite tensions and we live in the negative tension brane. The
hierarchy problem could be solved in this setup by the exponentially
changing metric along the extra dimension.

It has been shown firstly by Binetroy \emph{et al} [3] that the
Randall-Sundrum $2$-brane model in the absence of any source
(including a cosmological constant) in the bulk, leads to a
phenomenologically unacceptable cosmological evolution. However,
when there is a cosmological constant in the bulk, the usual
Friedmann equations are recovered at low energy but in this case
there is a strict constrain between the energy density of matter on
the branes which is undesirable phenomenologically [4]. Later, it
has been shown that the correct cosmological evolution equations are
recovered if one stabilizes the separation between the two branes
[5]. To do this end, one should introduce a massive scalar field
called Radion to stabilize the inter-brane separation. The simplest
mechanism for radius stabilization has been proposed by Goldberger
and wise (GW) [6]. They introduced a bulk scalar field with
different vacuum expectation values (VEVs) on the two branes. After
the stabilization, there is no constraint on the matter density on
the branes. Cosmological dynamics and phenomenology of the model are
crucially dependant on the mass and coupling of the Radion field.
One could achieve the desired result if the mass of the Radion is of
the order of $TeV$ and its couplings to Standard Model (SM) fields
is \emph{O}($TeV^{-1}$). It has been shown in reference [7] that in
the GW mechanism, these conditions are indeed satisfied. Reference
[8] provides a thorough analysis of the Radion dynamics and RS I
cosmology.

One faces the issue of Radion stabilization only in the RS I and ADD
solutions. After proposing the 2-brane model, Randall and Sundrum
introduced a new scenario [9] (RS II), in which  the extra dimension
is infinite in size, which effectively  means moving the negative
tension brane of RS I scenario to infinity. There is only a single
(positive tension) brane and no Radion is needed; its wave function
diverges away from the brane. Because of this simplicity, much of
the early works on brane cosmology have focused on the RS II
solution (see [10] and references therein).

In recent years, cosmologists have shown renewed interest in
studying the cosmological implications of the RS I model [11]. The
reason for this is that one can include the quantum gravity effects
more readily in this scenario: Kaluza-Klein excitations of the bulk
graviton modes are supposed to be equivalent to bound states of a
strongly coupled, nearly conformal field theory residing on the TeV
brane [12]. At temperatures above the TeV scale, the TeV brane is
assumed hidden behind a horizon [13] which is formed by a black hole
in the bulk [14]. In fact, at high temperatures there are two
possible scenarios. The first is that the TeV brane simply does not
exist at early times; the horizon shields the region where the TeV
brane might exist. As the temperature drops and the horizon recedes,
it eventually uncloaks the TeV brane, and the true SM degrees of
freedom emerge. An alternative possibility with similar physical
consequences is that a TeV brane exists at the scale associated with
the temperature of the theory. Only when the temperature drops to
appropriately low scale will the brane settle at its true minimum,
analogously to the behavior of other moduli in the early universe.
It has been noted that the emergence of the TeV brane from the
horizon may occur around the same time as the electroweak phase
transition [15]. Ref. [16] provided an alternative picture of a
first order phase transition which leads to the appearance of the
TeV brane at this epoch.

Although standard Big Bang cosmology has been very successful in
explaining the evolution of the universe as we see today; but
despite all of its successes there exist some problems unsolved in
this framework. The most notable of these are the flatness problem,
the observed low density of monopoles and the horizon problem
[17,18]. Theories with inflation (a period in the early universe
characterized by an exponential expansion of the scale factor) offer
solutions to these problems and in fact are the only ones to do so.
Inflation, however, suffers from its own set of problems. Primary
amongst these, is the generation of a correct (scalar) potential
that would drive inflation [19], and equally important the mechanism
of a non-contrived exit of the universe from an inflationary phase.
This graceful exit problem has been plaguing both cosmologists and
string theorists with no simple solution in sight. Extension of the
inflation paradigm to braneworld scenarios have provided a variety
of new ideas in the spirit of the particle physics and cosmology.

With these preliminaries, in this paper, we investigate the chaotic
inflation scenario in the Randall-Sundrum 2-brane (RS I) model. We
assume that the size of the extra dimension is stabilized using
appropriate Radion potential. The modified Friedmann equations for
the branes are used to obtain the evolution of the universe in the
inflation era. We adopt two relatively different approaches in which
in the first approach, the stabilizing field potential directly
appears in the Friedmann equation and the resulting scenario is
effectively a two-field inflation. In the second approach, the
stabilization mechanism is considered in the context of a warp
factor so that there is just one field present that plays the roles
of both inflaton and stabilizer. These two approaches are compared
and by analyzing the parameter space of the model we investigate
possible realization of the graceful exit in this setup. We study
constraints imposed on the model parameters from recent
observations.

\section{The model}
Consider a RS model with two branes located at $y=0$ and $y=1/2$. We
suppose that the metric has the form
\begin{eqnarray}
ds^2&=&n^{2}(y,t) dt^2-a^{2}(y,t) (dx_1^2+dx_2^2+dx_3^2)-b^{2}(y,t)
dy^2,
\nonumber \\
&\equiv& \tilde{g}_{AB}(x,y) dx^A dx^B.
\end{eqnarray}
where $y$ is the coordinate of the extra dimension and
$A,B=0,1,2,3,5$. Components of the Einstein tensor for this metric
are [3]
\begin{eqnarray}
&& G_{00}=3\left[ \left(\frac{\dot{a}}{a}\right)^2
+\frac{\dot{a}\dot{b}}{ab} -\frac{n^2}{b^2} \left( \frac{a''}{a}
+\left(\frac{a'}{a}\right)^2 -\frac{a'b'}{ab}\right) \right], \nonumber \\
&& G_{ii}=\frac{a^2}{b^2} \left[ \left(\frac{a'}{a}\right)^2+ 2
\frac{a'}{a}\frac{n'}{n} -\frac{b'n'}{bn}-2 \frac{b'a'}{ba}
+2\frac{a''}{a}+\frac{n''}{n}\right]+ \frac{a^2}{n^2}
\left[-\left(\frac{\dot{a}}{a}\right)^2+2
\frac{\dot{a}}{a}\frac{\dot{n}}{n}-2\frac{\ddot{a}}{a}+\right.
\nonumber \\
&& \left.\frac{\dot{b}}{b} \left(
-2\frac{\dot{a}}{a}+\frac{\dot{n}}{n} \right) -\frac{\ddot{b}}{b}
\right],
\nonumber \\
&& G_{05}=3\left[ \frac{n'}{n}\frac{\dot{a}}{a}
+\frac{a'\dot{b}}{ab}-\frac{\dot{a}'}{a}\right],
\nonumber \\
&& G_{55}=3 \left[
\frac{a'}{a}\left(\frac{a'}{a}+\frac{n'}{n}\right) -\frac{b^2}{n^2}
\left(
\frac{\dot{a}}{a}\left(\frac{\dot{a}}{a}-\frac{\dot{n}}{n}\right)
+\frac{\ddot{a}}{a}\right) \right],
\end{eqnarray}
where a prime denotes differentiation with respect to $y$ while a
dot marks differentiation with respect to $t$, the cosmic time. We
assume that there exists a cosmological constant in the bulk so the
overall energy-momentum tensor has two parts. One from the bulk
cosmological constant of the form
 \begin{equation}
T_{ab}^{(bulk)}=\tilde{g}_{ab} \Lambda,
\end{equation}
 and the other
from the matter on the branes which has the following form
\begin{eqnarray}
T_{a}^{b\,\,(brane)}= &&\frac{\delta (y)}{b} {\rm diag}\
\Big(V_*+\rho_*\, ,\, V_*-p_*\,,\, V_*-p_*\,,\, V_*-p_*\, ,\, 0\Big)
+\nonumber \\
&&\frac{\delta (y-\frac{1}{2})} {b} {\rm diag} \ \Big(-V+\rho\, ,\,
-V-p\, ,\, -V-p\, ,\, -V-p\, ,\, 0\Big),
\end{eqnarray}
where $V_*$ is the (positive) tension (or equivalently the
cosmological constant) of the  brane located at $y=0$,\, $\rho_*$
and $p_*$ are the density and pressure of the matter situated  on
the positive tension brane (with an equation of state of the form
$p_*=w_* \rho_*$ ) and $\rho$ and $p$ are the density and pressure
of the matter on the negative tension brane (the brane which we live
in). Both sets of densities and pressures are measured with respect
to $\tilde{g}$. In the absence of any energy-momentum sources on the
branes, that is, where \, $\rho,\,\, p,\,\, \rho_{*},\,\, p_{*}\,\to
0$, one recovers the static Randall-Sundrum solution of the form
 \begin{equation}
 n(y)=a(y)=e^{-|y|m_0 b_0},
 \end{equation}
 where the relations between $\Lambda,\, V_*$,\, $V$  and $m_0$ are
given by
\begin{eqnarray}
&& V_*=\frac{6m_0}{\kappa^2}= -V, \nonumber \\
&& \Lambda =-\frac{6m^2 _0}{\kappa^2}.
\end{eqnarray}
The effective 4D Planck scale is then given by
\begin{equation} (8
\pi G_N)^{-1}= M_{Pl}^2 \equiv \frac{1-\Omega^2_0}{\kappa^2 m_0}
\hbox{,}
\end{equation}
where $\Omega_{0}$ is the present-day value of the warp factor and
is given by
\begin{equation}
\Omega_0 \equiv e^{-m_0b_0/2}
\end{equation}
The warp factor in general is $b$-dependent so we define
\begin{equation}
\Omega_{b}\equiv e^{-m_{0}b(t)|y|/2}
\end{equation}
where we measure it at $y=\frac{1}{2}$, at our brane. When
$b=b_0=constant$, then $\Omega_{b}=\Omega_{0}$.

Now we include the Radion in our equations. Without stabilization,
the Radion has no mass, so only its potential enters into the
action. We assume that the coefficients in the metric (1) have the
following forms
\begin{eqnarray}
a(y,t)&=&a(t) \Omega(y,b(t)) \left[1+
\delta \bar{a}(y,t) \right] \nonumber \\
n(y,t)&=&\Omega(y,b(t))\left[1+\delta \bar{n}(t,y) \right]
\nonumber \\
b(t,y)&=&b(t) \big[1+ \delta \bar{b}(y,t)\big] .
\end{eqnarray}
The modified Friedmann equation for the geometry given by the metric
(1) is [5]
\begin{equation}
\frac{\dot{a}^2}{a^2}+ (m_0 b) \frac{\Omega^2_b}{1-\Omega^2_b}
\frac{\dot{a}}{a} \frac{\dot{b}} {b} - \frac{(m_0 b)^2}{4}
\frac{\Omega^2 _b}{1- \Omega^2 _b} \frac{ \dot{b}^2}{b^2}=
\frac{\kappa^2 m_0} {3}\frac{1}{1-\Omega^2_b} \left(\rho_* +\rho
\Omega^4 _b +W_r(b)\right) + \epsilon^2
\end{equation}
where $\epsilon^2 = O\Big((\delta \bar{a} )^2,\, (\delta \bar{n}
)^2,\, (\delta \bar{b})^2\Big)$ and $W_r(b)$ is the Radion
potential. From the above Friedmann equation one can deduce that
unless there is a strict constraint between the matter densities on
the two branes, the universe evolves in a very unconventional manner
which is not supported by present day observations [3,5]. Also it is
very undesirable that the matter densities on the two branes should
obey such a constraints.

In order to avoid the above mentioned problems, we should stabilize
the radius. This can be done by assuming that the Radion is massive.
An elegant mechanism for Radion stabilization has been proposed by
Goldberger and Wise. After the stabilization, the Friedmann
equations are [5]
 \begin{eqnarray}
\frac{\dot{\bar{a}}^2}{\bar{a}^2}&=& \frac{8 \pi G_N}{3} \left(
f^{4}(b)( \rho_* + \rho_{vis}) +  \frac{1}{4} \frac{3}{8 \pi G_N}
(m_0 b)^2 \left( \frac{\Omega_b}{1-\Omega^2_b}\right)^2
\frac{\dot{b}^2}{b^2}
+ \overline{W}_r(b) \right) \nonumber \\
&=&  \frac{8 \pi G_N}{3} \left( \frac{1}{2} \dot{\psi}^2+
\overline{W}_r(\psi) + f^{4}(b)( \rho_* + \rho_{vis}) \right) \hbox{
,}
\end{eqnarray}
and
\begin{equation}
\frac{\ddot{\bar{a}}}{\bar{a}}=-4\pi G_{N}\bigg(f^{4}(b)\Big[(
\rho_* + \rho_{vis})+3( p_* +
p_{vis})\Big]+2(\dot{\psi^{2}}-{\overline{W}_r(\psi)}\bigg).
\end{equation}
Here $\psi$ is the canonically normalized Radion and we performed a
conformal transformation of the metric as
\begin{eqnarray}
a(t) &=& f(b(\bar{t})) \hbox{ } \bar{a} (\bar{t}) \nonumber \\
d t &=& f(b(\bar{t}))  \hbox{ }d \bar{t}  \hbox{ , }
\end{eqnarray}
where the function $f(b)$ is defined by
\begin{equation}
f(b) = \left( \frac{1- \Omega^2 _{0}}{1-\Omega^2_b} \right)^{1/2}
\hbox{.}
\end{equation}
With the above definitions, the Radion potential in the absence of
any source is
\begin{equation}
\overline{W}_{r}(b)=  f^{4}(b) W_r(b),
\end{equation}
and the canonically normalized Radion is given by
\begin{equation}
m_0 b(t) = \sqrt{\frac{2}{3}} \frac{\psi(t)}{\Omega_{0}M_{pl}}
(1-\Omega^2_0) \sim \sqrt{\frac{2}{3}}
\frac{\psi(t)}{\Omega_{0}M_{pl}} \hbox{ .}
 \end{equation}
The Friedmann equations (12) and (13) are derived by neglecting the
back-reaction of the Radion mass on the background metric and the
wave function of the Radion. However, it has been shown in reference
[8] that including these considerations will not change the results.
It should also be noted that equations (12) and (13) are represented
in the Einstein frame so in the low energy limit with a stabilized
Radion ($\dot{b}=0$), the universe evolves as in usual cosmology in
this frame. With the definition (14), it is obvious that up to an
small fluctuation, the evolution of the universe is also the same in
the original frame.
\section{The Goldberger-Wise Stabilization Mechanism}
The model for Radion stabilization which we use here is the one
proposed by Goldberger and Wise [6]. In their model, a bulk scalar
field which also contains two potentials on TeV and Planck branes,
plays the role of stabilizer. The whole action for this model is
\begin{equation}
S=\int
d^{5}x\sqrt{-\tilde{G}}\bigg(-\frac{1}{2\kappa_{5}^{2}}R-\Lambda+
\tilde{G}^{AB}\delta_{A}\Phi\delta_{B}\Phi-U_{bulk}(\Phi)\bigg)+S_{TeV}+S_{Pl},
\end{equation}
where the action on the branes are given by
\begin{eqnarray}
\nonumber S_{Pl}&=&\int \sqrt{-g}d^{4}x\Big(L_{m0}-U_{0}(\Phi)\Big)\\
S_{Tev}&=&\int \sqrt{-g}d^{4}x\Big(L_{m1/2}-U_{1/2}(\Phi)\Big).
\end{eqnarray}
Here $\tilde{G}_{AB}$ is the 5D metric given by equation (1) and the
potentials in the bulk and in the brane are given by
\begin{eqnarray}
\nonumber U_{bulk}(\Phi)&=&\frac{1}{2}m^{2}_{s}\Phi^{2}\\
U_{i}(\Phi)&=&\lambda_{i}(\Phi^{2}-v^{2}_{i})^{2}
\end{eqnarray}
where $i=0,\, \frac{1}{2}$ and $v_{i}$'s have the dimension of
$(mass)^{3/2}$ and could be different on the two branes. With this
action, Goldberger and Wise find the effective 4D potential for the
Radion as
\begin{equation}
W_r(b)=4 m_0 e^{-2 m_0 b} \left(v_{1/2}-v_0 e^{-\epsilon m_0 b/2}
\right)^2 \Big(1+ \frac{\epsilon}{4}\Big) - \epsilon m_0 v_{1/2}
e^{-(4+ \epsilon)m_0 b/2} \left( 2 v_{1/2} -v_0 e^{-\epsilon m_0
b/2}\right)
\end{equation}
where $\epsilon$ is defined by
\begin{equation}
\epsilon=\frac{4m^{2}_{s}}{m_{0}}.
\end{equation}
Substituting from equation (17) for the canonically normalized
Radion, we get
$$W_r(\psi)=4 m_0 e^{-2 (\sqrt{\frac{2}{3}}
\frac{\psi}{\Omega_{0}M_{pl}})} \left(v_{1/2}-v_0 e^{-\epsilon
(\sqrt{\frac{2}{3}} \frac{\psi}{\Omega_{0}M_{pl}})/2} \right)^2 (1+
\frac{\epsilon}{4} )$$
\begin{equation}
\quad\quad\quad\,\, - \epsilon m_0 v_{1/2} e^{-(4+
\epsilon)(\sqrt{\frac{2}{3}} \frac{\psi}{\Omega_{0}M_{pl}})/2}
\left( 2 v_{1/2} -v_0 e^{-\epsilon (\sqrt{\frac{2}{3}}
\frac{\psi}{\Omega_{0}M_{pl}})/2}\right).
\end{equation}
Figure $1$ shows the shape of the Golberger-wise stabilization
potential in our setup as given by equation (23).
\begin{figure}[htp]
\begin{center}\includegraphics{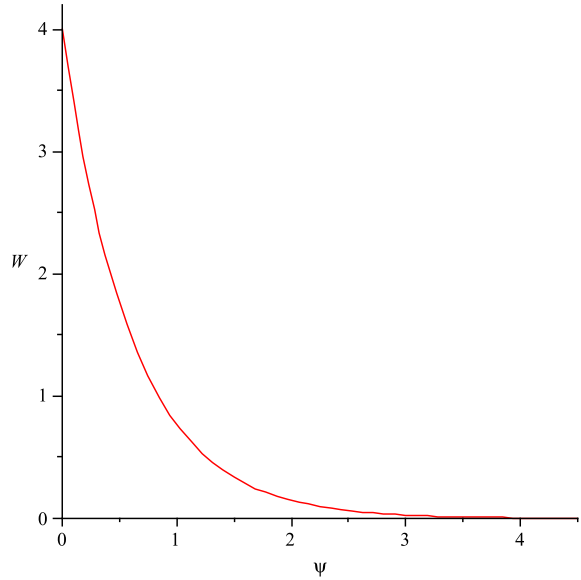} \vspace{7cm}
\end{center}
\caption{\small {Stabilization potential in Goldberger-Wise
mechanism as given by equation (23).}}
\end{figure}

\section{Chaotic inflation in the RS I model}
Now we have the tools to investigate the inflation in the
Randall-Sundrum 2-brane scenario. Our starting points are the
modified Friedmann equations (12) and (13). We assume that the
inflation takes place in the TeV brane so we set $\rho_{*}=0$.

We assume also that there is an inflaton field $\phi$, with a
potential $V(\phi)$ on the TeV brane where its energy density and
pressure are given as
\begin{equation}
\rho=\frac{1}{2}\dot{\phi}^{2}+V
\end{equation}
and
\begin{equation}
 p=\frac{1}{2}\dot{\phi}^{2}-V
\end{equation}
respectively. To have an inflationary era, there should be a phase
of accelerated expansion in the early stages of the universe
evolution in which $\ddot{\bar{a}}>0$. From equation (13) this means
that
\begin{equation}
f^{4}(b)\Big( \rho_{vis}+3
p_{vis}\Big)+2\Big(\dot{\psi^{2}}-{\overline{W}_r(\psi)}\Big)<0
\end{equation}
Substituting (24) and (25) into (26) we get
\begin{equation}
f^{4}(b)\bigg[2(\dot{\phi^{2}}-V)\bigg]+2\bigg(\dot{\psi^{2}}-
{\overline{W}_r(\psi)}\bigg)<0\,,
\end{equation}
which is the condition for realization of inflation in our setup.
The Klein-Gordon equation governing the evolution of the inflaton
field is as usual
\begin{equation}
\ddot{\phi}+3H\dot{\phi}+V'(\phi)=0.
\end{equation}
In the standard Friedmann cosmology, the energy condition required
for realization of inflation yields \,
$\frac{1}{2}\dot{\phi^{2}}<V(\phi)$\,  which means that the
potential energy of the inflaton dominates over its kinetic energy.
Imposing the slow-roll approximation \emph{i.e.}
\begin{equation}
\frac{1}{2}\dot{\phi^{2}}\ll V(\phi)
\end{equation}
and
\begin{equation}
\ddot{\phi}\ll 3H\dot{\phi},
\end{equation}
on our model, the Hubble parameter for a model universe where the
only matter is an inflaton field on the TeV brane, is given by
equation (12) as
\begin{equation}
H^{2}=  \frac{8 \pi G_N}{3} \left[ \frac{1}{2} \dot{\psi}^2+
\overline{W}_r(\psi) + f^{4}(b)\Big(
\frac{1}{2}\dot{\phi}^{2}+V(\phi)\Big) \right] \hbox{ .}
\end{equation}
In the slow-roll approximation this becomes
\begin{equation}
H^{2}\simeq  \frac{8 \pi G_N}{3} \left[ \frac{1}{2} \dot{\psi}^2+
\overline{W}_r(\psi) + f^{4}(b)\Big(V(\phi)\Big) \right] \hbox{ .}
\end{equation}
Also from Klein-Gordon equation one finds
\begin{equation}
\dot{\phi}\simeq-\frac{V'(\phi)}{3H}.
\end{equation}
We assume that potential energy of the Radion field dominates its
kinetic energy so that the first term on the right hand side of
equation (32) can be neglected. Now, the slow-roll parameters are
defined as
\begin{eqnarray}
\nonumber\varepsilon&\equiv&\frac{M^{2}_{pl}
}{16\pi}\bigg(\frac{W'+f V'}{W+fV}\bigg)^{2}\\
\eta&\equiv&\frac{M^{2}_{pl} }{16\pi}\bigg(\frac{W''+f
V''}{W+fV}\bigg),
\end{eqnarray}
where a prime denotes differentiation with respect to the argument
in each case. We note that these definitions of the slow-roll
parameters are very similar to the two-field inflation. So,
inflation in our setup is effectively a two-field inflation. In
other words, incorporation of the role played by Radion in this
inflation scenario turns it to an effective two-field inflation. The
slow-roll approximations are valid when
\begin{eqnarray}
\nonumber\varepsilon&\ll& 1\\|\eta|&\ll&1\,.
\end{eqnarray}
The inflationary phase ends when $\varepsilon$ and $|\eta|$ grow of
order unity. The amount of inflation is described by the number of
e-folds which is defined as
\begin{equation}
N\equiv\ln(\frac{a_{f}}{a_{i}})=\int^{t_{f}}_{t_{i}}Hdt.
\end{equation}
Substituting for $H$ from equation (32) we find
\begin{equation}
N=\int^{\psi_{f}}_{\psi_{i}}\int^{\phi_{f}}_{\phi_{i}}\bigg[\frac{8
\pi G_N}{3} \left( \frac{1}{2} \dot{\psi}^2+ \overline{W}_r(\psi) +
f^{4}(b)\Big(V(\phi)\Big) \right)\bigg ]^{1/2}d\phi d\psi.
\end{equation}
It is obvious that the number of e-folds in our setup is larger than
the standard case due to incorporation of the Radion potential which
is positive as figure $1$ shows. Figures $2$ and $3$ show the
graceful exit from the inflationary phase. The condition for
graceful exit is $\varepsilon=1$, however it is possible essentially
to fulfill the condition $|\eta|=1$ before fulfilling
$\varepsilon=1$ and in this case inflation terminates when the
condition $|\eta|=1$ is fulfilled.

\begin{figure}[htp]
\begin{center}\includegraphics{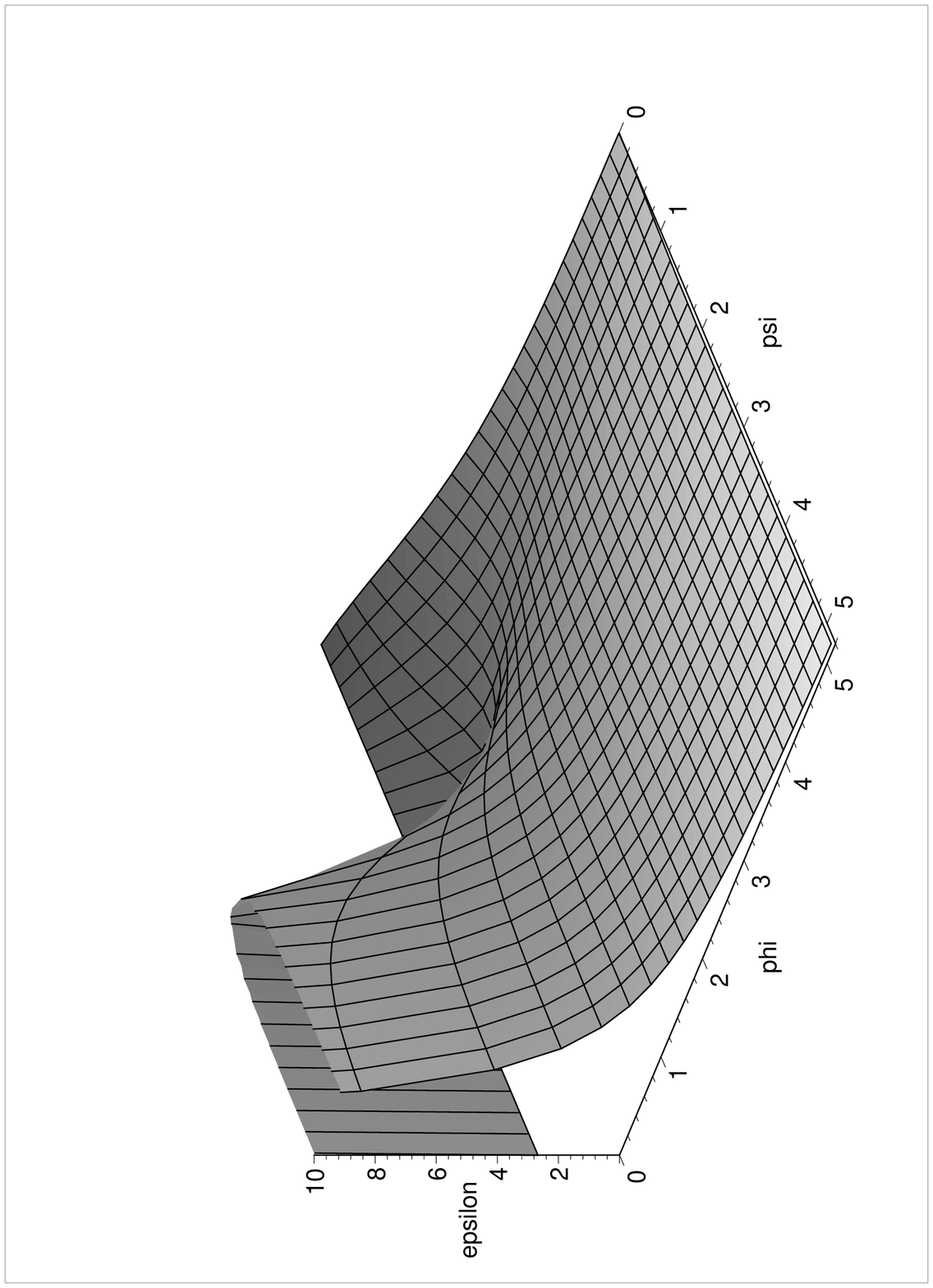} \vspace{0cm}
\end{center}
\vspace{6cm}
 \caption{\small {Graceful exit from inflationary phase. $\varepsilon$ versus $\phi/M_{pl}$ and $\psi/M_{pl}$}}
\end{figure}

\begin{figure}[htp]
\begin{center}\includegraphics{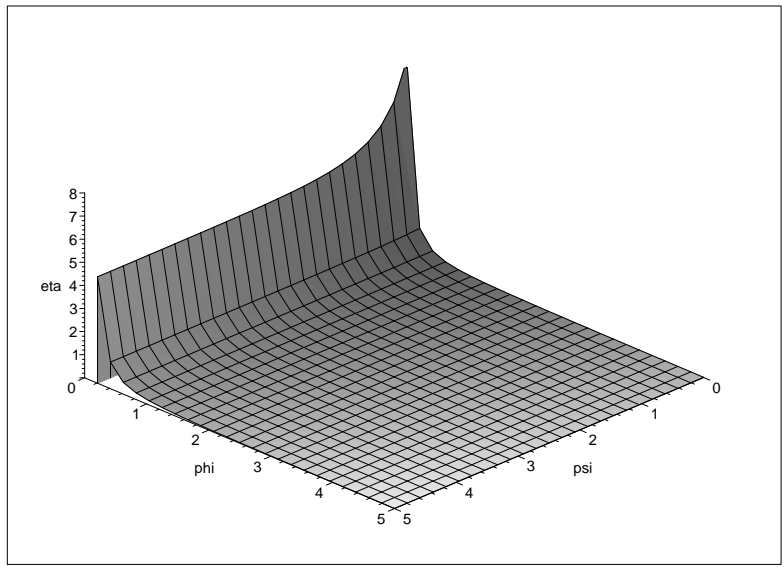} \vspace{3cm}
\end{center}
\vspace{6cm}
 \caption{\small {Graceful exit from inflationary phase. $\eta$ versus $\phi/M_{pl}$ and $\psi/M_{pl}$.}}
\end{figure}
\newpage
\section{Perturbations}
Density perturbations due to quantum mechanical fluctuations of the
inflaton field, have important observational consequences for the
present epoch of the universe because, it is believed that these
perturbations are the source of large scale structure formation in
the universe.\\
The fluctuations of the inflaton field are related to the curvature
perturbations on comoving hypersurfaces by the following relation
 \begin{equation}
\emph{R}=\frac{H\delta\phi}{\dot{\phi}}
 \end{equation}
In the slow-roll limit the field fluctuation at the horizon crossing
(or Hubble radius) is given by
 \begin{equation}
 \delta\phi=\frac{H}{2\pi}
 \end{equation}
 This result is independent of the geometry
and holds for a massless scalar field in de Sitter spacetime
independently of the gravitational field equations. The amplitude of
scalar perturbations is related to the density perturbation by the
following relation
\begin{equation}
A^{2}_{s}=4\frac{\emph{R}^{2}}{25}
\end{equation}
In our model, using equations (31),(32) and (38), this gives
\begin{equation}
A^{2}_{s}\simeq \frac{2048}{675}\bigg(\frac{\pi
(G_{N})^{3}}{V'^{2}}\bigg)\left( \frac{1}{2} \dot{\psi}^2+
\overline{W}_r(\psi) + f^{4}(b)V(\phi) \right)^{3} \hbox{ ,}
\end{equation}
which is calculated at the Hubble crossing. One of the important
parameters which is useful to constraint the inflationary models
with observations is the spectral index, $n_{s}$. In the slow-roll
limit, $n_{s}$ is given by
\begin{equation}
n_{s}-1\equiv\frac{d\ln A^{2}_{s}}{d\ln k}=-6\varepsilon+2\eta
\end{equation}
where $k$ is the comoving wavenumber. Recent WMAP five year result
[20] combined with SDSS and SNIa data shows that $n_{s}\simeq
0.960$. We plot the value of $n_{s}$ versus  $\phi$ and $\psi$
fields at the end of inflation in figure $4$. Figure $5$ shows the
values of $\phi$ and $\psi$ to match the observed value of $n_{s}$
by WMAP. We use this figure to extract the values of $\phi_{end}$
and $\psi_{end}$ and then using equation (37) we find that the
number of e-folds in our model is $N\simeq 65-105$ which is slightly
larger than the standard case which is expected as has been
discussed after equation (37). It is also obvious from figure $5$
that with suitable value of $\psi$ the value of inflaton field
$\phi$ could be below the planck scale thus avoiding the so called
$\eta$-problem.
\begin{figure}[htp]
\begin{center}\includegraphics{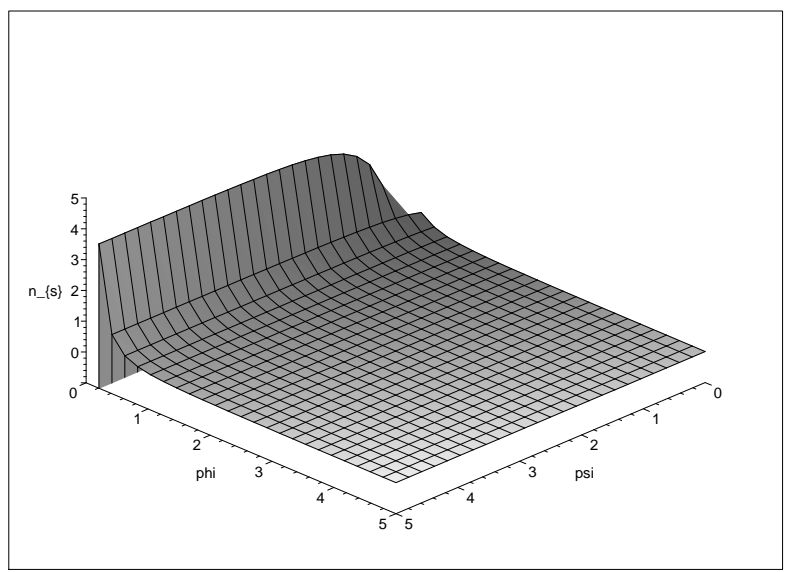} \vspace{0cm}
\end{center}
\vspace{6cm}
 \caption{\small {Value of $n_{s}$ for different values of $\phi$ and $\psi$ fields.}}
\end{figure}

\begin{figure}[htp]
\begin{center}\includegraphics{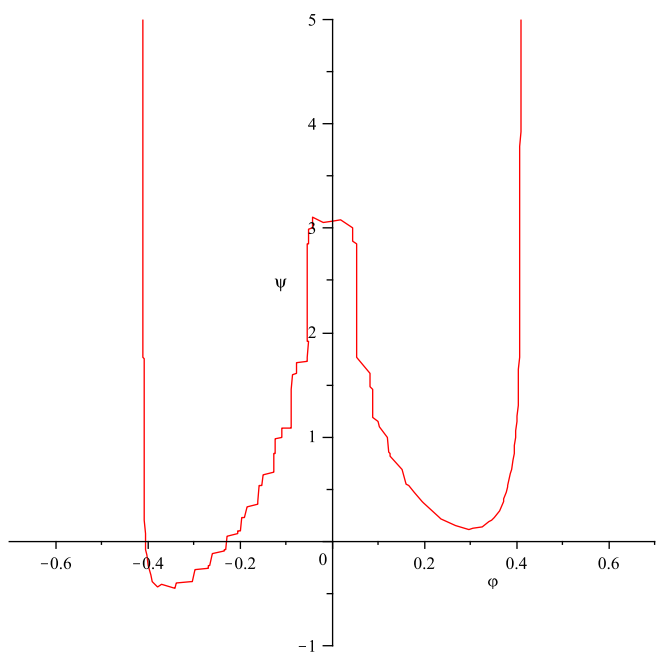} \vspace{1.5cm}
\end{center}
\vspace{7cm}
 \caption{\small {Acceptable values of $\phi$ and $\psi$ to match the observed value of
 $n_{s}$ by combined  WMAP5+SDSS+SNIa dataset. }}
\end{figure}
The amplitude of tensorial perturbations at the Hubble crossing is
\begin{equation}
A^{2}_{T}=\frac{4}{25\pi}\bigg(\frac{H}{M_{pl}}\bigg)^{4}.
\end{equation}
In the slow-roll approximation using equations (31) this gives
\begin{equation}
A^{2}_{T}\simeq \frac{32 \pi (G_{N})^{2}}{75} \left( \frac{1}{2}
\dot{\psi}^2+ \overline{W}_r(\psi) + f^{4}(b)V(\phi) \right) \hbox{
.}
\end{equation}
Figures $6$ and $7$ shows the amplitude of scalar and tensorial
perturbations for our model. These figures show that our model is
favored by the recent observations from combined WMAP5+SDSS+SNIa
dataset.
\begin{figure}[htp]
\begin{center}
\includegraphics{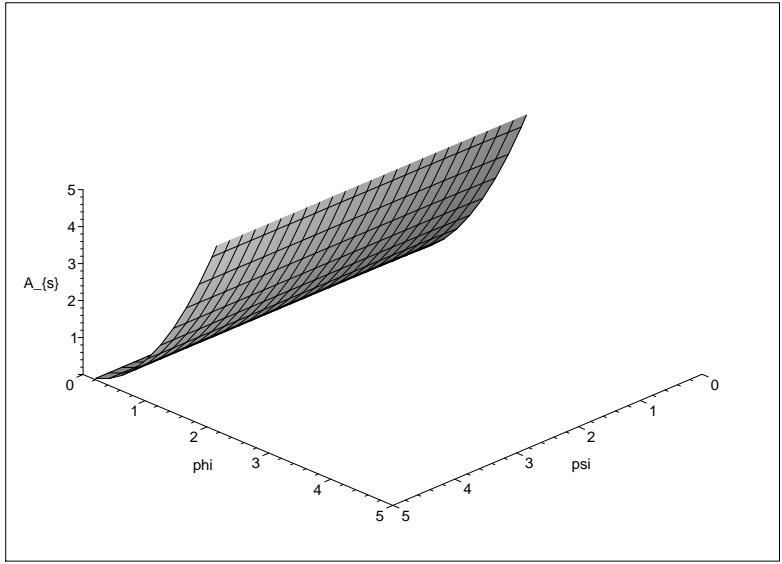} \vspace{5cm}
\end{center}
\vspace{3cm}
 \caption{\small {Amplitude of scalar  perturbations vs $\phi$ and $\psi$ fields . }}
\end{figure}
\begin{figure}[htp]
\begin{center}
\includegraphics{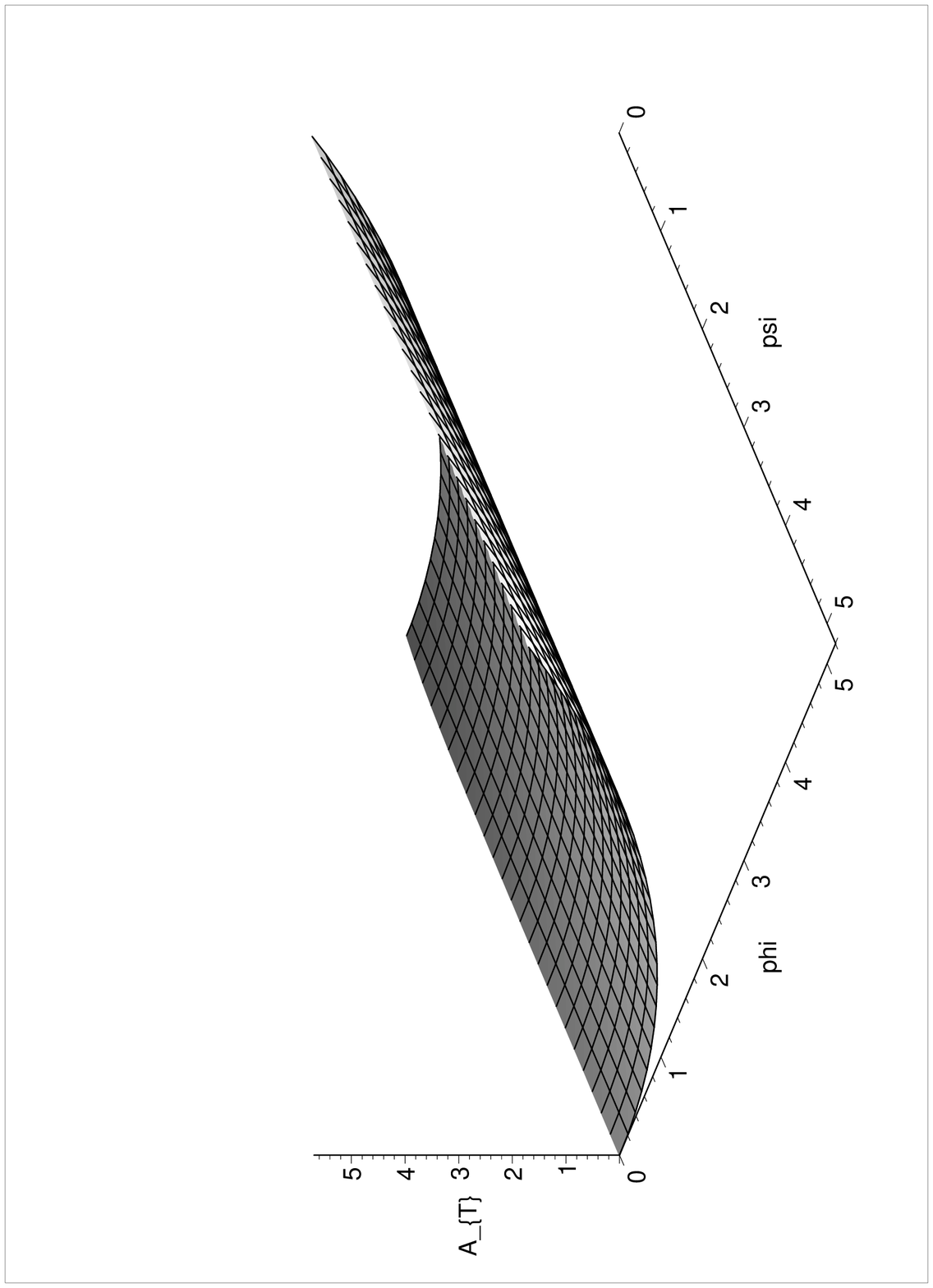} \vspace{5cm}
\end{center}
\vspace{1cm}
 \caption{\small {Amplitude of  tensorial perturbations vs $\phi$ and $\psi$ fields  }}
\end{figure}
The tensorial spectral index is given by
\begin{equation}
n_{T}\equiv \frac{d\ln A^{2}_{T}}{d\ln k}=-2\varepsilon.
\end{equation}
The relative amplitude of tensorial and scalar spectrums is
therefore
\begin{equation}
\frac{A^{2}_{T}}{A^{2}_{R}}=0.140625\frac{V'^{2}}{G_{N}}\left(
\frac{1}{2} \dot{\psi}^2+ \overline{W}_r(\psi) + f^{4}(b)(V(\phi))
\right)^{-2} \hbox{ .}
\end{equation}
Figure $8$ shows the relative amplitude of tensorial and scalar
perturbations for our model
\begin{figure}[htp]
\begin{center}\includegraphics{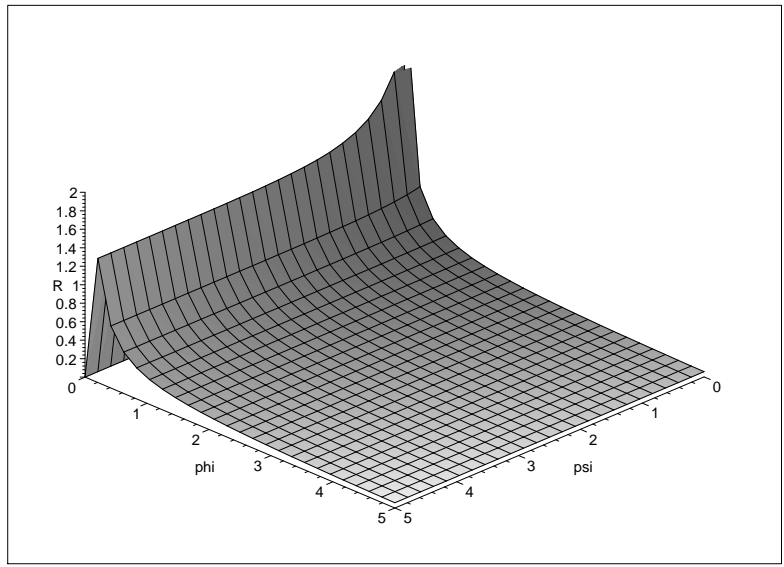} \vspace{1.5cm}
\end{center}
\vspace{6cm}
 \caption{\small {Relative Amplitude of tensorial and scalar perturbations vs $\phi$ and $\psi$ fields }}
\end{figure}
\newpage
\section{An alternative approach}
An alternative representation of the Randall-Sundrum model mentioned
in section 2, was presented by Cline and Firouzjahi [21]. Using
action (18) for a RS I model with radion stabilization, they made a
perturbative expansion of the coefficients of the metric as
$$
n(t,y)=n_{0}(y)+\delta n(t,y)\,\,\,\,\,\,a(t,y)=a_{0}(y)+\delta
a(t,y)
$$
\begin{equation}
b(t,y)=b_{0}+\delta
b(t,y)\,\,\,\,\,\,\,\quad,\quad\Phi(t,y)=n_{0}(y)+\delta \Phi(t,y)
\end{equation}
The Klein-Gordon equation for the scalar field, $\Phi$, for this
action is
\begin{equation}
\delta_{t}\bigg(\frac{1}{n}ba^{3}\dot{\Phi}\bigg)
-\delta_{y}\bigg(\frac{1}{b}a^{3}n\Phi'\bigg)+ba^{3}n\bigg[U'+U'_{0}\delta(by)+U'_{1/2}\delta(b(y-1/2))\bigg]=0
\end{equation}
Inserting the metric in the form of (1) in Einstein equations with
stress-energy tensor as (4), the solutions in zeroth order in
perturbations are
\begin{equation}
\Phi_{0}(y)\simeq v_{0}e^{-\epsilon m_{0}b_{0}y/2};\quad\quad
a_{0}(y)\simeq
m_{0}b_{0}y+\frac{\kappa^{2}}{12}v^{2}_{0}\big(e^{-\epsilon
m_{0}b_{0}y/2}-1\big)
\end{equation}
Where $\epsilon$ and $v _{0}$ are given in section $3$. The
Friedmann equations are
\begin{eqnarray}
\nonumber\bigg(\frac{\dot{a}_{0}}{a_{0}}\bigg)^{2}&=&\frac{8\pi
G}{3}\bigg(\rho_{*}+\Omega^{4}\rho\bigg)\\
\bigg(\frac{\dot{a}_{0}}{a_{0}}\bigg)^{2}-\frac{\ddot{a}_{0}}{a_{0}}&=&4\pi
G \bigg(\rho_{*}+p_{*}+\Omega^{4}(\rho+p)\bigg).
\end{eqnarray}
Here the warp factor, $\Omega$, is defined as
\begin{equation}
\Omega\equiv e^{-a_{0}(1/2)}
\end{equation}
where $a_{0}(y)$ is given by (49). This equation is valid before the
stabilization when $b=b_{0}$ but in the inflation era, $b_{0}$ in
the equation (49) should be replaced by $b$ which is a function of
time, so we have
\begin{equation}
\Omega(\Phi)=\exp\Bigg[\frac{8}{m_{0}}\ln\frac{\Phi}{v_{0}}+\frac{\kappa^{2}}{3}v^{2}_{0}\Big(e^{-\epsilon
\frac{1}{m_{0}}\ln\frac{\Phi}{m_{0}}}-1\Big)\Bigg].
\end{equation}
Now we turn our attention to inflation and we assume again that the
only matter existed in the universe is an inflaton field whose
energy density and pressure are given by (24) and (25) respectively.
We note that the inflaton field plays the role of stabilizer in this
setup. Substituting (24) and (49) into Friedmann equation (50), we
get
\begin{equation}
H^2=\frac{8\pi
G}{3}\Bigg\{\exp\bigg[\frac{8}{m_{0}}\ln\frac{\Phi}{v_{0}}+\frac{\kappa^{2}}{3}v^{2}_{0}\Big(e^{-\epsilon
\frac{1}{m_{0}}\ln\frac{\Phi}{m_{0}}}-1\Big)\bigg]\Big(\frac{1}{2}\dot{\Phi}^{2}+U(\Phi)\Big)\Bigg\}.
\end{equation}
Choosing $U(\Phi)$ to be
\begin{equation}
U(\Phi)=\frac{1}{2}m^{2}_{s}\Phi^{2}
\end{equation}
and imposing the slow roll conditions \emph{i.e}
\begin{equation}
\frac{1}{2}\dot{\Phi^{2}}\ll U(\Phi),
\end{equation}
the slow-roll parameters are
\begin{equation}
\varepsilon=\frac{m^{2}_{pl}}{4\pi}\Bigg(\frac{\Omega'(\Phi)U(\Phi)+\Omega(\Phi)U'(\Phi)}{2\Omega(\Phi)U(\Phi)}\Bigg)^{2}
\end{equation}
and
\begin{equation}
\eta=\frac{m^{2}_{pl}}{4\pi}\frac{\Big[\Omega''(\Phi)U(\Phi)+\Omega(\Phi)U''(\Phi)+2\Omega'(\Phi)U'(\Phi)\Big]
\Omega(\Phi)U(\Phi)+\Big[\Omega'(\Phi)U(\Phi)+\Omega(\Phi)U'(\Phi)\Big]^{2}}{2\Omega(\Phi)U(\Phi)},
\end{equation}
where a prime denotes differentiation with respect to $\Phi$ and
$\Omega(\Phi)$ and $U(\Phi)$ are given by equations (52) and (54)
respectively. Figures $9a$ and $9b$ show the value of $\varepsilon$
and $\eta$ respectively. It can be seen from these figures that the
inflationary phase ends when $\Phi_{f}\simeq (0.11-0.2 ) M_{pl}$
where $\Phi_{f}$ is the value of $\Phi$ at the end of the inflation.
These values of $\Phi$ lead to the number of e-folds $N\simeq
62-95$. Also one should note that the value of the $\Phi$-field at
the end of the inflation is below the planck scale so the
$\eta$-problem [22] will not occur in this model. Figure $10$ shows
the value of $n_{s}$ in this case. The observed value of $n_{s}$
\emph{i.e} $n_{s}\simeq 0.96$ occurs at $\Phi\simeq 0.16 M_{pl}$
which is consistent with previous result. Figure $11$ shows the
tensorial spectral index. For the value of $\Phi\simeq 0.16 M_{pl}$
, we find from this figure that $n_{T}\simeq-0.2$ where lies in the
acceptable range supported by observations. There exist a
consistency relation between the tensorial spectral index,$n_{T}$
and tensor-to-scalar ratio , $r$ as[17]
\begin{equation}
n_{T}=-\frac{r}{8}
\end{equation}
so we find that the tensor-to-scalar ratio in our model is $r=0.16$.
The constraint on $r$ given in WMAP5 is $r<0.22$ so we can see that
our result is acceptable.

\begin{figure}[htp]
\begin{center}
\includegraphics{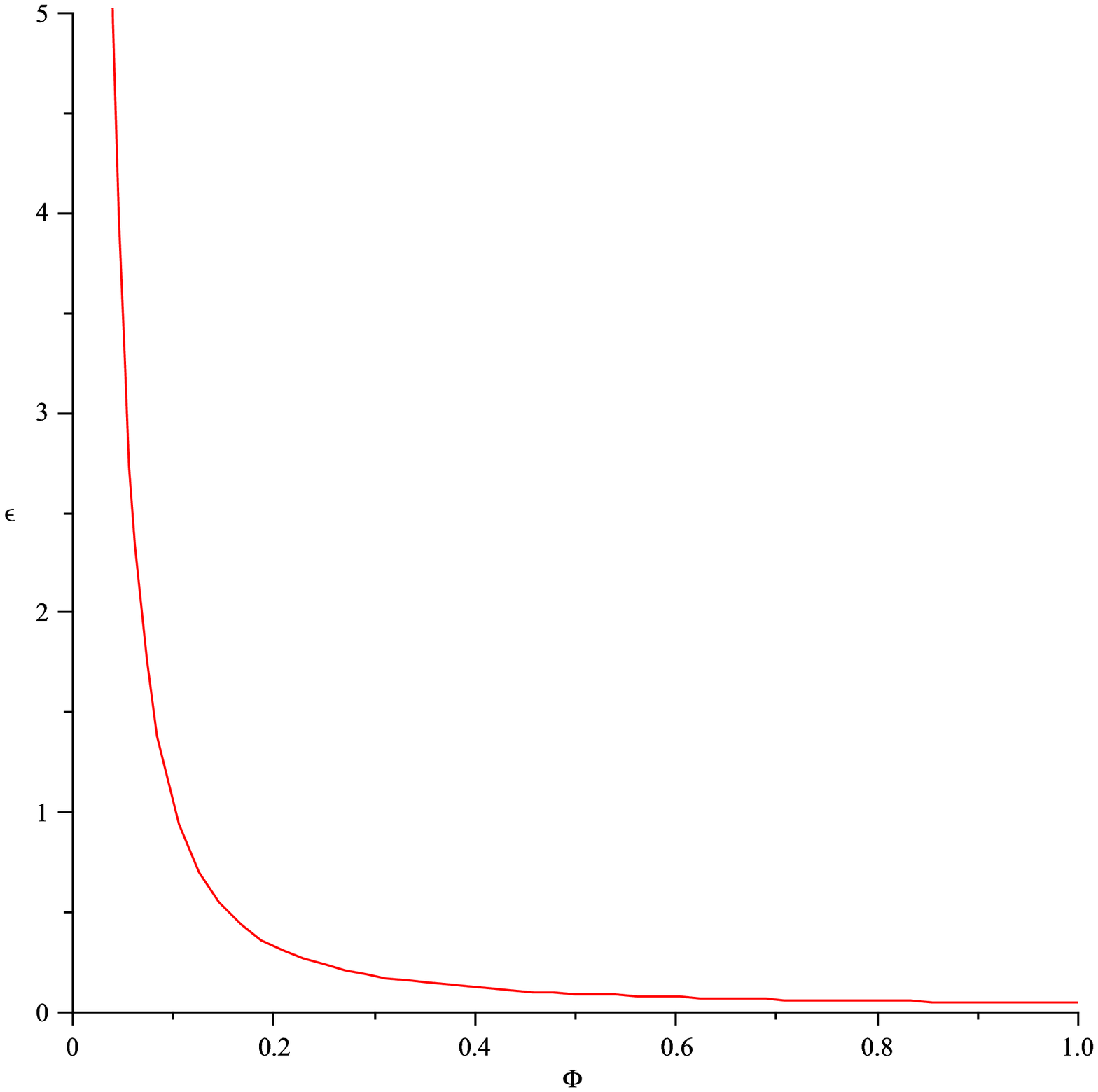} \vspace{5cm}\includegraphics{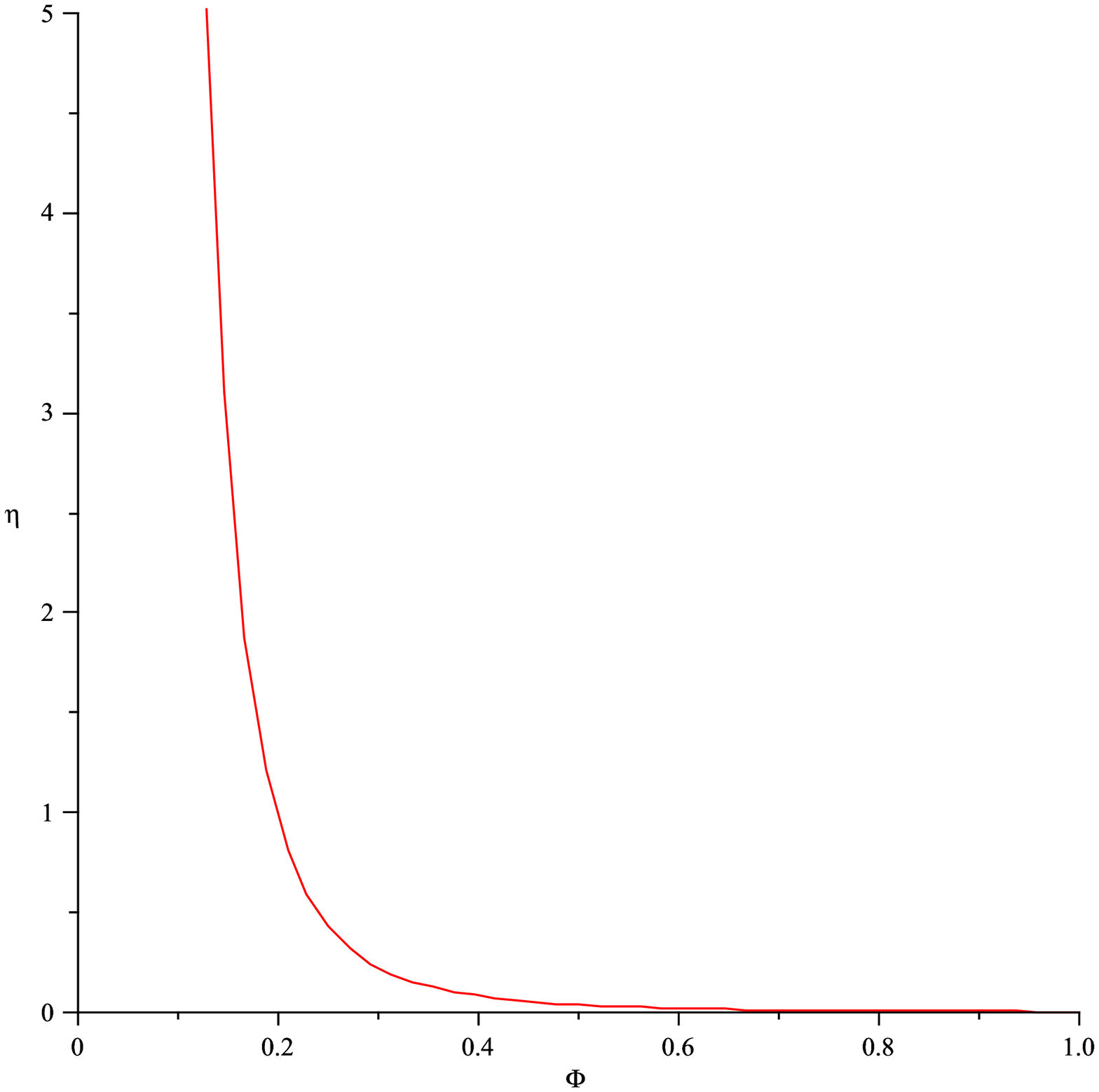}\vspace{2 cm}
\end{center}
\vspace{1cm}
 \caption{\small {Graceful exit from the inflationary phase: a) $\varepsilon$ versus $\Phi/M_{pl}$. b) $\eta$ versus $\Phi/M_{pl}$.  }}
\end{figure}
\begin{figure}[htp]
\begin{center}\includegraphics{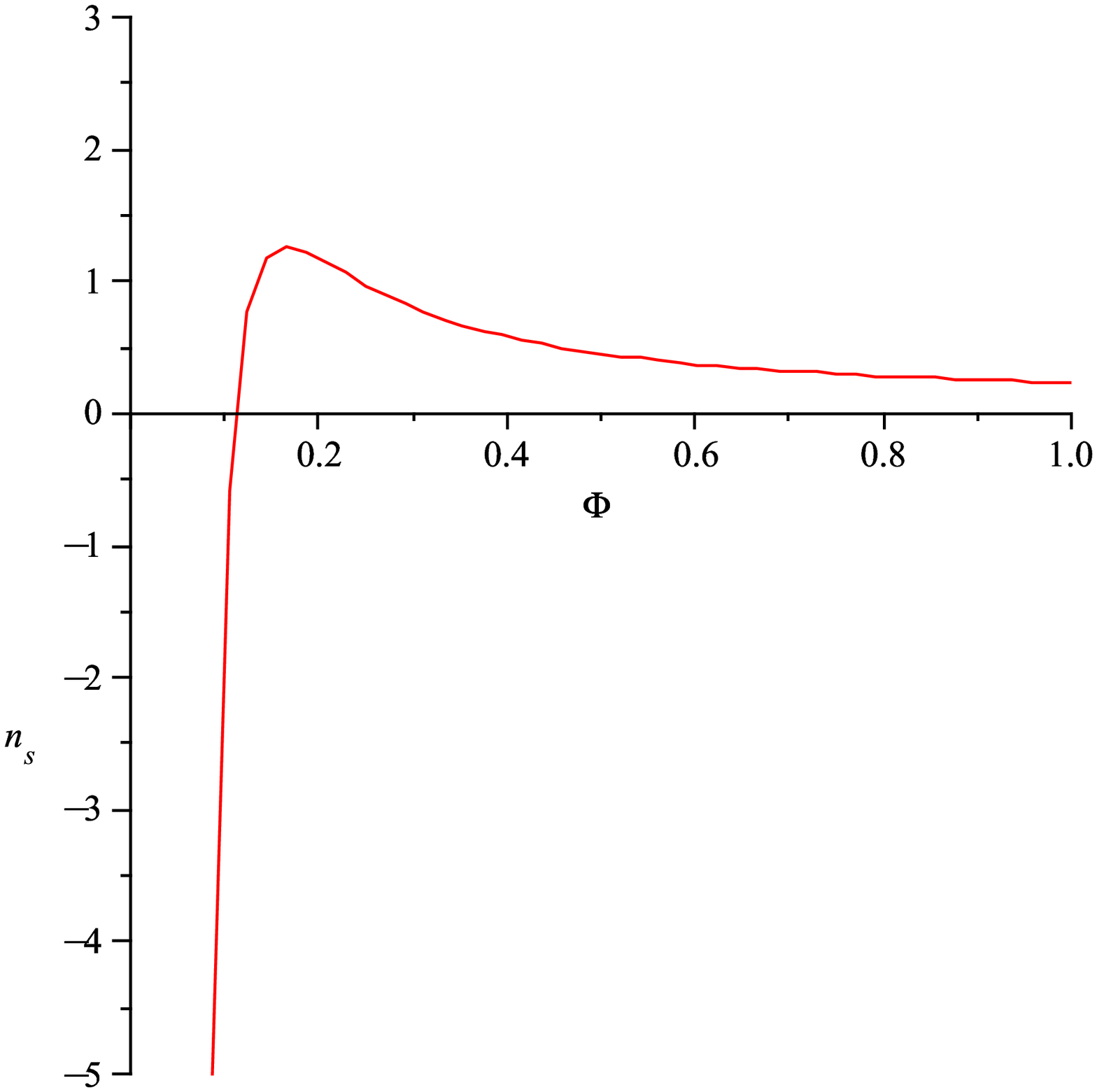} \vspace{1.5cm}
\end{center}
\vspace{6cm}
 \caption{\small {The value of $n_{s}$ versus $\Phi/M_{pl}$}}
\end{figure}
\begin{figure}[htp]
\begin{center}\includegraphics{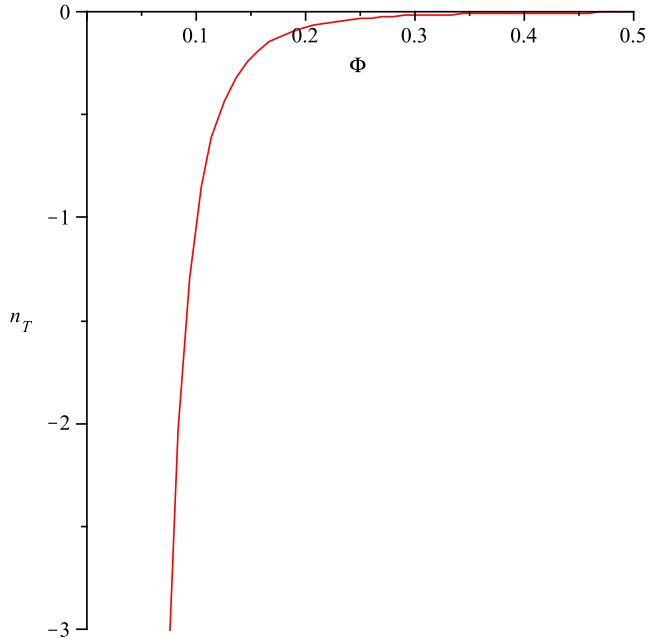} \vspace{1.5cm}
\end{center}
\vspace{6cm}
 \caption{\small {The value of $n_{T}$ versus $\Phi/M_{pl}$}}
\end{figure}
\newpage
\section{Conclusion}
Incorporating an inflationary phase in the Randall-Sundrum 2-brane
scenario is a challenging problem because one should consider the
issue of inter-brane stabilization. In this paper, we tried to
construct a suitable model that matches with existing observational
data of the combined WMAP5+SDSS+SNIa dataset. We did this using two
different approaches: First we included the stabilization potential
and Radion kinetic term directly in the Friedmann equations. The
stabilization mechanism that we used is the Goldberger-Wise
mechanism in which a bulk scalar field plays the role of the
stabilizer. Assuming that inflation takes place in TeV brane, this
model is effectively a two-field inflation scenario. One is the
usual inflaton field and the other is a new field corresponding to
the inter-brane separation. The inflation parameters \emph{i.e}
number of e-folds, scalar spectral index and tensor-to-scalar ratio
in this model are consistent with observational data from combined
WMAP5+SDSS+SNIa dataset with suitable values of the Radion field
$\psi$. Also the value of inflaton field at the end of the
inflation, $\phi_{end}$ is below the planck scale and therefore
$\eta$ problem will not occur in this setup.

The second approach is based on the assumption that the
stabilization field , $\Phi$ , also plays the role of the inflaton
field. Using the explicit relation for the warp factor given in
reference [21], we calculated the inflation parameters in this
model. Our results in this case is also consistent with
observational data from WMAP5+SDSS+SNIa.\\

{\bf Acknowledgement}\\
We are indebted to Professor Hassan Firouzjahi for his invaluable
comments on this work.


\begin{thebibliography}{21}
\bibitem{1}
N. Arkani-Hamed, S. Dimopoulos and G. Dvali, Phys. Lett. B {\bf 429}
(1998) 263, [arXiv:9803315]; I. Antoniadis, N. Arkani-Hamed, S.
Dimopoulos and G. Dvali, Phys. Lett. B {\bf 436} (1998) 257,
[arXiv:9804398]
\bibitem{2}
L. Randall and R. Sundrum, Phys. Rev. Lett. {\bf 83} (1999) 3370,
[arXiv:9905221]
\bibitem{3}
P. Binetruy, C. Deffayet and D. Langlois, Nucl. Phys. B {\bf 565}
(2000) 269, [arXiv:9905012]; P. Binetruy, C. Deffayet, U. Ellwanger
and D. Langlois, Phys. Lett. B {\bf 477} (2000) 285, [arXiv:9910219]
\bibitem{4}
Shiromizu, K. Maeda and M. Sasaki, Phys. Rev. D {\bf 62} (2000)
024021, [arXiv:9910076]; J. M. Cline, C. Grojean and G. Servant,
Phys. Rev. Lett. {\bf 83} (1999) 4245, [arXiv:9906523]; E. E.
Flanagan, S. H. Tye and I. Wasserman , [arXiv:9910498]
\bibitem{5}
C. Cs´aki, M. Graesser, L. Randall and J. Terning, Phys. Rev. D {\bf
62} (2000) 045015, [arXiv:9911406]
\bibitem{6}
W. D. Goldberger and M. B. Wise, Phys. Rev. Lett. {\bf 83} (1999)
4922, [arXiv:9907447]; W. D. Goldberger and M. B. Wise, Phys. Rev. D
{\bf 60} (1999) 107505, [arXiv:9907218]
\bibitem{7}
W. D. Goldberger and M. B. Wise, Phys. Lett. B {\bf 475} (2000) 275,
[arXiv:9911457]
\bibitem{8}
C. Csaki, M. L. Graesser and G. D. Kribs, Phys. Rev. D {\bf 63}
(2001) 0650020, [arXiv:0008151]
\bibitem{9}
L. Randall and R. Sundrum, Phys. Rev. Lett. {\bf 83} (1999) 4690,
[arXiv:9906064]
\bibitem{10}
D. Langlois, Prog. Theor. Phys. Suppl. {\bf 148} (2003) 181,
 [arXiv:0209261]
\bibitem{11}
J. M. Cline, J. Vinet, JHEP {\bf 0202} (2002) 042, [arXiv:0201041]
\bibitem{12}
S. S. Gubser, Phys. Rev. D {\bf 63} (2001) 084017, [arXiv:99912001]
\bibitem{13}
N. Arkani-Hamed, M. Porrati and L. J. Randall, JHEP {\bf 0108}
(2001) 017, [arXiv:0012148]
\bibitem{14}
A. Hebecker and J. March-Russell, Nucl. Phys. B {\bf 608} (2001)
375, [arXiv:0103214]
\bibitem{15}
P. Creminelli, A. Nicolis and R. Rattazzi, JHEP {\bf 0203} (2002)
051 [arXiv:0107174]
\bibitem{16}
J. M. Cline and H. Firouzjahi, Phys. Rev. D {\bf 64} (2001) 023505,
[arXiv:0005235]
\bibitem{17}
A. R. Liddle and D. H. Lyth, {\it Cosmological Inflation and
Large-Scale Structure}, Cambridge University Press, 2000.
\bibitem{18}
R. H. Brandenberger, [arXiv:0509099]; J. E. Lidsey {\it et al}, Rev.
Mod. Phys. {\bf69} (1997) 373, [arXiv:9508078]
\bibitem{19}
A. R. Liddle and D. H. Lyth, Phys. Rep {\bf 231} (1993) 1,
 [arXiv:9303019]
\bibitem{20}
E. Kumatsu \emph{et.al}, Astrophys. J. Suppl. {\bf 180} (2009) 330, 
[arXiv:0803.0547]
\bibitem{21}
J. M. Cline and H. Firouzjahi, Phys.Lett. B \textbf{495} (2000)
271-276, [arXiv:0008185]
\bibitem{22}
D. H. Lyth and A. Riotto, Phys. Rep. \textbf{314}, 1 (1999)
[arXiv:9807278]
\end{thebibliography}
\end{document}